\documentclass[a4paper]{article}
\RequirePackage[english]{babel}
\RequirePackage[latin1]{inputenc}
\RequirePackage[T1]{fontenc}
\RequirePackage{mathrsfs}
\RequirePackage{amsmath}
\RequirePackage{amssymb}
\RequirePackage{amsbsy}
\RequirePackage{bm}
\begin{document}
\title{\bf{The Spin-Torsion coupling and Causality\\ for the Standard Model}}
\author{Luca Fabbri\\ 
\footnotesize INFN \& Gruppo Teorici, Dipartimento di Fisica dell'Universit\`{a} di Bologna\\
\footnotesize GPP, D\'{e}partement de Physique de l'Universit\'{e} de Montr\'{e}al}
\date{}
\maketitle
\begin{abstract}
We study the influence of the spin-torsion coupling prescription and we prove the causal propagation for the fields of the standard model.
\end{abstract}
\section*{Introduction}
In the late 1960s, Salam and Weinberg built a theory of fields characterized by the $U(1) \times SU(2)_{L}$ local gauge symmetry among the massless fields in which the mechanism of symmetry breaking generates the masses of the interacting fields (\cite{w}): because the fields involved in the Weinberg and Salam model get their mass while still interacting with one another then the issue of acausality as described by Velo and Zwanziger may arise (\cite{v-z}); moreover additional mass corrections and effective interactions due to the spin-torsion coupling postulated by Kibble and Sciama may emerge (\cite{k} and \cite{s}). In this paper we will consider the SW standard model, applying the VZ general methods to show that the causal propagation of all fields is preserved, and studying the KS spin-torsion coupling to see how it may influence the whole system of fields.
\section{The Weinberg-Salam model}
First, we recall the structure of the Weinberg-Salam standard model \cite{w}.

We start by considering the Lagrangian in which the matter contribution is given in terms of fermion fields, for which the interaction with torsion and metric is given by the most general covariant derivative implemented in the vierbein form plus the complex phase transformation $U(1)$, and these fermion fields will be considered to be a couple of fermion fields verifying the condition of masslessness so that we will be able to separate the left- and right-handed projections and we will further be able to re-arrange the two left-handed and the two right-handed projections into two doublets, each of which could consequently mix according to an additional transformation $SU(2)$; we will first postulate that the mixing will occur only for the left-handed doublet according to the $SU(2)_{L}$ transformation, leaving the right-handed doublet without possibility to mix so that for simplicity we will consider it to be a right-handed singlet: the fermionic content can be generalized by considering bosons represented by complex scalar fields arranged in one doublet transforming under the same transformation. Now the transformation is $U(1) \times SU(2)_{L}$ which in general acts upon the fermions and the scalars independently, although we can and will postulate such a correlation between the coupling constants by defining
\begin{eqnarray}
&R'=e^{-i\alpha}R \ \ \ \ 
L'=e^{-\frac{i}{2}\left(\vec{\sigma} \cdot \vec{\theta}+\mathbb{I}\alpha\right)}L\\
&\phi'=e^{-\frac{i}{2}\left(\vec{\sigma} \cdot \vec{\theta}-\mathbb{I}\alpha\right)}\phi
\end{eqnarray}
and by considering the transformation to be local so that gauge fields $\vec{A}_{\mu}$ and $B_{\mu}$ are introduced transforming as
\begin{eqnarray}
\vec{\sigma}\cdot\vec{A}'_{\mu}=e^{-\frac{i}{2}\vec{\sigma} \cdot \vec{\theta}}
\left[\vec{\sigma}\cdot \left(\vec{A}_{\mu}-\frac{1}{g}\partial_{\mu}\vec{\theta}\right)\right]
e^{\frac{i}{2}\vec{\sigma} \cdot \vec{\theta}} \ \ \ \ \ \ \ \ \ \ \ \ 
B'_{\mu}=B_{\mu}-\frac{1}{g'}\partial_{\mu}\alpha
\end{eqnarray}
so that
\begin{eqnarray}
&D_{\mu}R=\nabla_{\mu}R-ig'B_{\mu}R\ \ \ \ \ \ \ \ \ \ 
D_{\mu}L=\nabla_{\mu}L
-\frac{i}{2}\left(g\vec{\sigma}\cdot\vec{A}_{\mu}+g'\mathbb{I}B_{\mu}\right)L\\
&D_{\mu}\phi=\nabla_{\mu}\phi
-\frac{i}{2}\left(g\vec{\sigma}\cdot\vec{A}_{\mu}-g'\mathbb{I}B_{\mu}\right)\phi
\end{eqnarray}
are the gauge covariant derivatives and
\begin{eqnarray}
\vec{A}_{\mu\nu}=\partial_{\mu}\vec{A}_{\nu}-\partial_{\nu}\vec{A}_{\mu}
+g\vec{A}_{\mu}\times\vec{A}_{\nu}\ \ \ \ \ \ \ \ \ \ \ \ 
B_{\mu\nu}=\partial_{\mu}B_{\nu}-\partial_{\nu}B_{\mu}
\end{eqnarray}
are the gauge curvatures, whose covariant derivatives are 
\begin{eqnarray}
D_{\rho}\vec{A}^{\rho\mu}=\nabla_{\rho}\vec{A}^{\rho\mu}+g\vec{A}_{\rho}\times\vec{A}^{\rho\mu}
\ \ \ \ \ \ \ \ \ \ \ \ \ \ \ \ D_{\rho}B^{\rho\mu}=\nabla_{\rho}B^{\rho\mu}
\end{eqnarray}
where all derivatives have been defined in terms of the ordinary derivatives.

Finally the Lagrangian 
\begin{eqnarray}
\nonumber
&\mathscr{L}=
\frac{i}{2}\left(\overline{R}\gamma^{\mu}D_{\mu}R-D_{\mu}\overline{R}\gamma^{\mu}R\right)
+\frac{i}{2}\left(\overline{L}\gamma^{\mu}D_{\mu}L-D_{\mu}\overline{L}\gamma^{\mu}L\right)+\\
\nonumber
&+D_{\mu}\phi^{\dagger}D^{\mu}\phi+\lambda^{2}\left(v^{2}\phi^{2}-\frac{1}{2}\phi^{4}\right)
-G_{Y}\left(\overline{R}\phi^{\dagger}L+\overline{L}\phi R\right)-\\
&-\frac{1}{4}A^{2}-\frac{1}{4}B^{2}+G
\end{eqnarray}
is written in terms of the $G_{Y}$, $v^{2}$ and $\lambda^{2}$ parameters, and it is clearly the most general invariant Lagrangian possible; the variation of this Lagrangian yields the field equations given by
\begin{eqnarray}
&i\gamma^{\mu}D_{\mu}R-G_{Y}\phi^{\dagger}L=0\\
&i\gamma^{\mu}D_{\mu}L-G_{Y}\phi R=0
\end{eqnarray}
for the fermions with
\begin{eqnarray}
D^{2}\phi+\lambda^{2}\left(\phi^{2}-v^{2}\right)\phi+G_{Y}\overline{R}L=0
\end{eqnarray}
for the scalar and with
\begin{eqnarray}
&D_{\rho}\vec{A}^{\rho\mu}
-\frac{ig}{2}\left(D^{\mu}\phi^{\dagger}\vec{\sigma}\phi
-\phi^{\dagger}\vec{\sigma}D^{\mu}\phi\right)
+\frac{g}{2}\overline{L}\gamma^{\mu}\vec{\sigma}L=0\\
&D_{\rho}B^{\rho\mu}+\frac{ig'}{2}\left(D^{\mu}\phi^{\dagger}\phi-\phi^{\dagger}D^{\mu}\phi\right)
+\frac{g'}{2}\overline{L}\gamma^{\mu}L+g'\overline{R}\gamma^{\mu}R=0\\
\nonumber
&G_{\alpha\mu}-\frac{1}{2}g_{\alpha\mu}G
+\frac{1}{2}\left(\frac{1}{4}g_{\alpha\mu}B^{2}
-B_{\mu\rho}B_{\alpha}^{\phantom{\alpha}\rho}\right)+
\frac{1}{2}\left(\frac{1}{4}g_{\alpha\mu}A^{2}
-\vec{A}_{\mu\rho}\cdot\vec{A}_{\alpha}^{\phantom{\alpha}\rho}\right)+\\
\nonumber
&+\frac{1}{2}\left[D_{\mu}\phi^{\dagger}D_{\alpha}\phi+D_{\alpha}\phi^{\dagger}D_{\mu}\phi
-g_{\alpha\mu}D_{\rho}\phi^{\dagger}D^{\rho}\phi
-g_{\alpha\mu}\lambda^{2}\left(v^{2}\phi^{2}-\frac{1}{2}\phi^{4}\right)\right]+\\
&+\frac{i}{4}
\left[\left(\overline{L}\gamma_{\alpha}D_{\mu}L-D_{\mu}\overline{L}\gamma_{\alpha}L\right)
+\left(\overline{R}\gamma_{\alpha}D_{\mu}R-D_{\mu}\overline{R}\gamma_{\alpha}R\right)\right]=0\\
&Q_{\mu\alpha\beta}
-\frac{i}{4}\left(\overline{L}\{\gamma_{\mu},\sigma_{\alpha\beta}\}L
+\overline{R}\{\gamma_{\mu},\sigma_{\alpha\beta}\}R\right)=0
\end{eqnarray}
for all interactions, all of these field equations being $U(1) \times SU(2)_{L}$ covariant field equations for massless fields.

For such a Lagrangian or system of field equations, the potential is given by
\begin{eqnarray}
V=\frac{1}{2}\phi^{4}-v^{2}\phi^{2}
\end{eqnarray}
whose stationary vacuum $\phi^{2}=0$ is invariant but unstable and therefore it will move toward stability breaking the symmetry by spontaneously assuming the $\phi^{2}=v^{2}$ configuration; because of this spontaneous breaking of the symmetry, it is without loss of generality that we can always choose the special gauge called unitary gauge with respect to which we have
\begin{eqnarray}
\phi=\left(\begin{tabular}{c}
$0$\\ $v+H$
\end{tabular}\right)
\end{eqnarray}
as the configuration that breaks the symmetry and also
\begin{eqnarray}
&\vec{A}_{\mu}=\vec{M}_{\mu}\\ 
&B_{\mu}=N_{\mu}
\end{eqnarray}
which are vector fields but they are not gauge vector fields any longer and
\begin{eqnarray}
&L=\left(\begin{tabular}{c}
$\nu_{L}$\\ $e_{L}$
\end{tabular}\right)\\
&R=\left(e_{R}\right) 
\end{eqnarray}
for the fermions: these new fields can be collected as 
\begin{eqnarray}
e_{L}+e_{R}\equiv e
\end{eqnarray} 
and re-named as
\begin{eqnarray}
\nonumber
&\cos{\theta}N_{\mu}-\sin{\theta}M^{3}_{\mu}=A_{\mu}\\
&\sin{\theta}N_{\mu}+\cos{\theta}M^{3}_{\mu}=Z_{\mu}
\end{eqnarray}
and
\begin{eqnarray}
\frac{1}{\sqrt{2}}\left(M^{1}_{\mu}\pm iM^{2}_{\mu}\right)=W_{\mu}^{\pm}
\end{eqnarray}
where $g'=g\tan{\theta}$ in terms of the Weinberg angle, in order to have diagonal mass matrix. And finally we define the cosmological constant $v^{4}\lambda^{2}=4\Lambda$ along with the masses $vG_{Y}=m$ for the Dirac fermion and $\sqrt{2}\lambda v=m_{H}$ for the Higgs scalar field and also $gv=m_{W}\sqrt{2}=m_{Z}\sqrt{2}\cos{\theta}$ for the two Proca vector fields.

Now, because the symmetry has been only partially broken, there still is a residual $U(1)_{q}$ symmetry, and setting $g\sin{\theta}=q$, we have this $U(1)_{q}$ symmetry acting upon the fields as 
\begin{eqnarray}
&e'=e^{-i\beta}e\ \ \ \ \ \ \ \ \ \ \ \ \ \ \ \ \ \ \ \ \nu'=\nu\\
&H'=H\ \ \ \ \ \ \ \ \ Z_{\mu}'=Z_{\mu}\ \ \ \ \ \ \ \ W'^{\pm}_{\mu}=e^{\mp i\beta}W^{\pm}_{\mu} 
\end{eqnarray}
which is a $U(1)_{q}$ local transformation corresponding to the gauge field $A_{\mu}$ that transforms as
\begin{eqnarray}
A'_{\mu}=A_{\mu}-\frac{1}{q}\partial_{\mu}\beta
\end{eqnarray}
so that
\begin{eqnarray}
&D_{\mu}e=\nabla_{\mu}e-iqA_{\mu}e\ \ \ \ \ \ \ \ \ \ \ \ \ \ \ \ D_{\rho}\nu=\nabla_{\rho}\nu\\
&D_{\alpha}H=\nabla_{\alpha}H \ \ \ \ \ \ \ \ D_{\mu}Z^{\nu}=\nabla_{\mu}Z^{\nu}\ \ \ \ 
D_{\mu}W^{\pm}_{\nu}=\nabla_{\mu}W^{\pm}_{\nu}\mp iqA_{\mu}W^{\pm}_{\nu}
\end{eqnarray}
are the gauge covariant derivatives and
\begin{eqnarray}
F_{\mu\nu}=\partial_{\mu}A_{\nu}-\partial_{\nu}A_{\mu}
\end{eqnarray}
is the gauge curvature, whose covariant derivative is  
\begin{eqnarray}
D_{\rho}F_{\mu\nu}=\nabla_{\rho}F_{\mu\nu}
\end{eqnarray}
where all derivatives have been defined in terms of the ordinary derivatives.

At last we are able to write down all the field equations after the spontaneous breakdown of the gauge symmetry as
\begin{eqnarray}
\nonumber
&i\gamma^{\mu}D_{\mu}e+q\tan{\theta}\left[Z_{\mu}
-\frac{3\cot{\theta}}{32q}\left((2\sin{\theta})^{2}\overline{e}\gamma_{\mu}e
-2\overline{e}_{L}\gamma_{\mu}e_{L}+2\overline{\nu}\gamma_{\mu}\nu\right)\right]\gamma^{\mu}e-\\
\nonumber
&-\frac{g}{2\cos{\theta}}\left[Z_{\mu}
-\frac{3\cot{\theta}}{32q}\left((2\sin{\theta})^{2}\overline{e}\gamma_{\mu}e
-2\overline{e}_{L}\gamma_{\mu}e_{L}
+2\overline{\nu}\gamma_{\mu}\nu\right)\right]\gamma^{\mu}e_{L}+\\
\nonumber
&+\frac{g}{\sqrt{2}}\left[W_{\mu}^{+}
+\frac{3}{g4\sqrt{2}}\left(\frac{1-(2\sin{\theta})^{2}}{(2\sin{\theta})^{2}}\right)
\overline{\nu}\gamma_{\mu}e_{L}\right]\gamma^{\mu}\nu-\\
&-m\left[\frac{{v+H+\frac{3}{2G_{Y}}\left(\frac{\cos{\theta}}{2}\right)^{2}\overline{e}e}}{v}\right]e
-i\left[\frac{3}{2}\left(\frac{\cos{\theta}}{2}\right)^{2}i\overline{e}\gamma e\right]\gamma e=0
\label{electron}\\
\nonumber
&i\gamma^{\mu}D_{\mu}\nu
+\frac{g}{2\cos{\theta}}\left[Z_{\mu}
-\frac{3\cot{\theta}}{32q}\left((2\sin{\theta})^{2}\overline{e}\gamma_{\mu}e
-2\overline{e}_{L}\gamma_{\mu}e_{L}+2\overline{\nu}\gamma_{\mu}\nu\right)\right]\gamma^{\mu}\nu+\\
&+\frac{g}{\sqrt{2}}\left[W_{\mu}^{-}
+\frac{3}{g4\sqrt{2}}\left(\frac{1-(2\sin{\theta})^{2}}{(2\sin{\theta})^{2}}\right)
\overline{e}_{L}\gamma_{\mu}\nu\right]\gamma^{\mu}e_{L}=0
\label{neutrino}
\end{eqnarray}
for the fermion fields with the system
\begin{eqnarray}
&D^{2}H+m_{H}^{2}\left(\frac{H^{2}+3vH+2v^{2}}{2v^{2}}\right)H
-\left(\frac{m_{Z}^{2}Z^{2}}{2}+m_{W}^{2}W^{2}\right)\left(\frac{H+v}{v^{2}}\right)
+\frac{m}{2v}\overline{e}e=0
\label{Higgs}\\
&D_{\mu}Z^{\mu}+2Z^{\mu}\nabla_{\mu}\ln{\left(v+H\right)}
+\frac{m}{\sqrt{2}m_{Z}}\frac{i\overline{e}\gamma e}{\left(v+H\right)}=0
\label{constraintneutral}\\
&D_{\mu}W^{\mu+}-iq\tan{\theta}Z^{\mu}W_{\mu}^{+}+2W^{\mu+}\nabla_{\mu}\ln{\left(v+H\right)}
-\frac{m}{m_{W}}\frac{i\overline{\nu}\gamma e}{(v+H)}=0
\label{constraintcharged}
\end{eqnarray}
in which it is clear that for the doublet of complex scalars the $4$ field equations have been separated apart so to have $1$ field equation left for the real scalar while the other $3$ field equations have been transformed into $1$ real and $1$ complex conditions that are constraints for
\begin{eqnarray}
\nonumber
&D_{\mu}D^{[\mu}W^{\nu]+}+ig\cos{\theta}Z_{\mu}D^{[\mu}W^{\nu]+}
+ig\cos{\theta}D_{\mu}\left(Z^{\mu}W^{\nu+}-Z^{\nu}W^{\mu+}\right)-\\
\nonumber
&-iW_{\mu}^{+}\left(g\cos{\theta}D^{[\mu}Z^{\nu]}-qF^{\mu\nu}\right)
+\left(g\cos{\theta}\right)^{2}\left(W_{\mu}^{+}Z^{\mu}Z^{\nu}-W^{\nu+}Z^{2}\right)+\\
&+m_{W}^{2}W^{\nu+}\left(\frac{v+H}{v}\right)^{2}
+g^{2}\left(W^{\nu-}W_{\mu}^{+}W^{\mu+}-W^{2}W^{\nu+}\right)
+\frac{g}{\sqrt{2}}\overline{\nu}\gamma^{\nu}e_{L}=0
\label{weakcharged}\\
\nonumber
&D_{\mu}D^{[\mu}Z^{\nu]}+\left(g\cos{\theta}\right)^{2}
\left[\left(W^{\nu-}W^{\mu+}+W^{\mu-}W^{\nu+}\right)Z_{\mu}-2W^{2}Z^{\nu}\right]+\\
\nonumber
&+m_{Z}^{2}Z^{\nu}\left(\frac{v+H}{v}\right)^{2}
+ig\cos{\theta}\left(W_{\mu}^{+}D^{[\mu}W^{\nu]-}-W_{\mu}^{-}D^{[\mu}W^{\nu]+}\right)+\\
\nonumber
&+ig\cos{\theta}D_{\mu}\left(W^{\mu+}W^{\nu-}-W^{\nu+}W^{\mu-}\right)
+q\overline{e}\gamma^{\nu}e\tan{\theta}-\\
&-\frac{g}{2\cos{\theta}}
\left(\overline{e}_{L}\gamma^{\nu}e_{L}-\overline{\nu}\gamma^{\nu}\nu\right)=0
\label{weakneutral}
\end{eqnarray}
for the charged and neutral massive vectors leaving then
\begin{eqnarray}
\nonumber
&D_{\mu}F^{\mu\nu}-iq\left(W^{+}_{\mu}D^{[\mu}W^{\nu]-}-W^{-}_{\mu}D^{[\mu}W^{\nu]+}\right)-\\
\nonumber
&-iqD_{\mu}\left(W^{\mu+}W^{\nu-}-W^{\nu+}W^{\mu-}\right)-\\
&-qg\cos{\theta}\left[\left(W^{\nu-}W^{\mu+}+W^{\mu-}W^{\nu+}\right)Z_{\mu}-2W^{2}Z^{\nu}\right]
+q\overline{e}\gamma^{\nu}e=0
\label{electrodynamics}\\
\nonumber
&\left(R_{\alpha\mu}-\frac{1}{2}g_{\alpha\mu}R\right)+
\frac{1}{2}\left(\frac{1}{4}g_{\alpha\mu}F^{2}
-F_{\mu\rho}F_{\alpha}^{\phantom{\alpha}\rho}\right)+\\
\nonumber
&+\frac{1}{2}\left(\frac{1}{4}g_{\alpha\mu}D_{[\beta}Z_{\rho]}D^{[\beta}Z^{\rho]}
-g^{\rho\beta}D_{[\alpha}Z_{\beta]}D_{[\mu}Z_{\rho]}\right)+\\
\nonumber
&+\frac{1}{2}\left[\frac{1}{2}g_{\alpha\mu}D^{[\beta}W^{\rho]-}D_{[\beta}W_{\rho]}^{+}
-g^{\rho\beta}\left(D_{[\mu}W_{\rho]}^{+}D_{[\alpha}W^{-}_{\beta]}+
D_{[\mu}W_{\rho]}^{-}D_{[\alpha}W^{+}_{\beta]}\right)\right]+\\
\nonumber
&+\frac{i}{2}\left(qF_{\mu\rho}-g\cos{\theta}D_{[\mu}Z_{\rho]}\right)
\left(W^{+}_{\alpha}W^{\rho-}-W^{\rho+}W^{-}_{\alpha}\right)+\\
\nonumber
&+\frac{i}{2}\left(qF_{\alpha\rho}-g\cos{\theta}D_{[\alpha}Z_{\rho]}\right)
\left(W^{+}_{\mu}W^{\rho-}-W^{\rho+}W^{-}_{\mu}\right)+\\
\nonumber
&+\frac{i}{2}g_{\alpha\mu}
\left(g\cos{\theta}D^{[\beta}Z^{\rho]}-qF^{\beta\rho}\right)W^{-}_{\rho}W^{+}_{\beta}-\\
\nonumber
&-\frac{i}{2}g\cos{\theta}
\left(D_{[\mu}W^{+}_{\rho]}W^{-}_{\alpha}-D_{[\mu}W^{-}_{\rho]}W^{+}_{\alpha}\right)Z^{\rho}+\\
\nonumber
&+\frac{i}{2}g\cos{\theta}
\left(D_{[\mu}W^{+}_{\rho]}W^{\rho-}-D_{[\mu}W^{-}_{\rho]}W^{\rho+}\right)Z_{\alpha}-\\
\nonumber
&-\frac{i}{2}g\cos{\theta}
\left(D_{[\alpha}W^{+}_{\rho]}W^{-}_{\mu}-D_{[\alpha}W^{-}_{\rho]}W^{+}_{\mu}\right)Z^{\rho}+\\
\nonumber
&+\frac{i}{2}g\cos{\theta}
\left(D_{[\alpha}W^{+}_{\rho]}W^{\rho-}-D_{[\alpha}W^{-}_{\rho]}W^{\rho+}\right)Z_{\mu}+\\
\nonumber
&+\frac{i}{2}g_{\alpha\mu}g\cos{\theta}
\left(W^{-}_{\beta}D^{[\beta}W^{\rho]+}-W^{+}_{\beta}D^{[\beta}W^{\rho]-}\right)Z_{\rho}-\\
\nonumber
&-\frac{1}{2}\left(g\cos{\theta}\right)^{2}
\left[\left(W^{+}_{\mu}W^{-}_{\alpha}+W^{+}_{\alpha}W^{-}_{\mu}\right)Z^{2}
+2W^{2}Z_{\alpha}Z_{\mu}\right]+\\
\nonumber
&+\frac{1}{2}\left(W^{+}_{\rho}W^{-}_{\alpha}+W^{+}_{\alpha}W^{-}_{\rho}\right)
\left(g\cos{\theta}\right)^{2}Z_{\mu}Z^{\rho}+\\
\nonumber
&+\frac{1}{2}\left(W^{+}_{\mu}W^{-}_{\rho}+W^{+}_{\rho}W^{-}_{\mu}\right)
\left(g\cos{\theta}\right)^{2}Z_{\alpha}Z^{\rho}+\\
\nonumber
&+\frac{1}{2}g_{\alpha\mu}\left(g\cos{\theta}\right)^{2}
\left(W^{2}Z^{2}-W^{\beta+}W^{-}_{\rho}Z_{\beta}Z^{\rho}\right)-\\
\nonumber
&-\frac{1}{2}g^{2}\left(W^{+}_{\mu}W^{-}_{\alpha}+W^{+}_{\alpha}W^{-}_{\mu}\right)W^{2}+\\
\nonumber
&+\frac{1}{2}g^{2}\left(W^{-}_{\mu}W^{-}_{\alpha}W^{+}_{\rho}W^{\rho+}
+W^{+}_{\mu}W^{+}_{\alpha}W^{-}_{\rho}W^{\rho-}\right)+\\
\nonumber
&+\frac{1}{4}g_{\alpha\mu}g^{2}
\left[\left(W^{2}\right)^{2}-\left(W^{\beta-}W^{-}_{\beta}W^{+}_{\rho}W^{\rho+}\right)\right]+\\
\nonumber
&+\left(D_{\mu}H D_{\alpha}H-\frac{1}{2}g_{\alpha\mu}D_{\rho}H D^{\rho}H \right)
+\frac{1}{2}g_{\alpha\mu}m_{H}^{2}H^{2}\left(\frac{H+2v}{2v}\right)^{2}+\\
\nonumber
&+\frac{1}{2}m_{Z}^{2}\left(\frac{H+v}{v}\right)^{2}
\left(Z_{\alpha}Z_{\mu}-\frac{1}{2}g_{\alpha\mu}Z^{2}\right)+\\
\nonumber
&+\frac{1}{2}m_{W}^{2}\left(\frac{H+v}{v}\right)^{2}
\left(W^{+}_{\alpha}W^{-}_{\mu}+W^{+}_{\mu}W^{-}_{\alpha}-g_{\alpha\mu}W^{2}\right)+\\
\nonumber
&+\frac{i}{8}
\left[\left(\overline{\nu}\gamma_{\alpha}D_{\mu}\nu
-D_{\mu}\overline{\nu}\gamma_{\alpha}\nu\right)
+\left(\overline{\nu}\gamma_{\mu}D_{\alpha}\nu
-D_{\alpha}\overline{\nu}\gamma_{\mu}\nu\right)\right]+\\
\nonumber
&+\frac{i}{8}
\left[\left(\overline{e}\gamma_{\alpha}D_{\mu}e
-D_{\mu}\overline{e}\gamma_{\alpha}e\right)
+\left(\overline{e}\gamma_{\mu}D_{\alpha}e
-D_{\alpha}\overline{e}\gamma_{\mu}e\right)\right]+\\
\nonumber
&+\frac{1}{4}q\tan{\theta}
\left(\overline{e}\gamma_{\alpha}eZ_{\mu}+\overline{e}\gamma_{\mu}eZ_{\alpha}\right)-\\
\nonumber
&-\frac{g}{8\cos{\theta}}
\left(\overline{e}_{L}\gamma_{\alpha}e_{L}Z_{\mu}
+\overline{e}_{L}\gamma_{\mu}e_{L}Z_{\alpha}
-\overline{\nu}\gamma_{\alpha}\nu Z_{\mu}
-\overline{\nu}\gamma_{\mu}\nu Z_{\alpha}\right)+\\
\nonumber
&+\frac{g}{4\sqrt{2}}\left(\overline{e}_{L}\gamma_{\alpha}\nu W_{\mu}^{+}
+\overline{e}_{L}\gamma_{\mu}\nu W_{\alpha}^{+}
+\overline{\nu}\gamma_{\alpha}e_{L}W_{\mu}^{-}
+\overline{\nu}\gamma_{\mu}e_{L}W_{\alpha}^{-}\right)+\\
&+\frac{3}{64}g_{\alpha\mu}
\left(4\overline{\nu}\gamma_{\rho}\nu\overline{e}_{L}\gamma^{\rho}e_{L}
-2\overline{\nu}\gamma_{\rho}\nu\overline{e}\gamma^{\rho}e
-\overline{e}\gamma_{\rho}e\overline{e}\gamma^{\rho}e\right)
-g_{\alpha\mu}\Lambda=0
\label{gravity}
\end{eqnarray}
for all the massless interactions, and all of these field equations are still $U(1)_{q}$ covariant hermitian field equations but now for massive fields with a cosmological term: thus the spontaneous breakdown of the gauge symmetry has rearranged the $4$ field equations for the scalar for which $3$ of them have converted into $3$ constraints for $3$ gauge vector fields which have therefore acquired the additional degree of freedom they needed to become massive and as a consequence all Proca fields are now in interaction with one another.

\subsection{The Velo-Zwanziger problem}
Next, we remind to the reader the Velo-Zwanziger general methodology employed to discuss acausal propagation \cite{v-z}.

In general matter fields are classified according to the value of their spin, namely a given matter field of spin $s$ possesses $2s+1$ degrees of freedom: these will have to correspond to the $2s+1$ independent solutions of a system of differential equations that specify the highest-order time derivative for all components of the field, called system of matter field equations. However, it may happen that field equations are not enough to determine the correct rank of the solution, and thus restrictions need be imposed in terms of equations in which all components of the field have highest-order time derivatives that never occur, called constraints: these constraints can be imposed in two ways, either being implied by the field equations, or being assigned as subsidiary conditions that come along with the field equations themselves. Although the former procedure seems more elegant, whenever interactions are present it gives rise to the situation for which the interacting fields could let terms of the highest-order derivative appear in the equation that determines the propagation of the wave fronts, affecting the propagation of the wave fronts. The equation that determines the propagation of the wave fronts is obtained by considering in the field equations eventually modified by constraints only the terms of the highest-order derivative in which all derivatives will be formally replaced with the vector $n$ getting a matrix, called propagator or characteristic matrix, of which one has to require the singularity, giving an equation, called characteristic equation; its solutions $n$ represent the characteristic propagation of the wave fronts in such a way that if there is no time-like solutions then there is no acausal propagation of the wave fronts themselves. What happens is that for special types of field equations for massive fields the interactions let the commutator of covariant derivatives appear in the constraints, and thus within the field equations, modifying first the propagator, then the characteristic equation, finally the solutions, and therefore affecting the causality of the propagation.

This is a very general procedure that can be applied to any model: because in the SW standard model we have spontaneous breakdown of the gauge symmetry generating the mass of the massive fields and making all Proca fields interacting with each other, then we have the conditions for which causality issues as described by the Velo-Zwanziger problem can arise: so by considering for the massive vector fields the field equations (\ref{weakneutral}) and (\ref{weakcharged}) and plugging the constraints provided by (\ref{constraintneutral}) and (\ref{constraintcharged}) back into the correspondent field equation we see that both these field equations can equivalently be written as field equations in which the Laplacian is the only highest-order derivative term, the same is trivially true for the massive scalar field with field equations (\ref{Higgs}), nothing happens for the massless bosonic fields with field equations (\ref{electrodynamics}) and (\ref{gravity}), and obviously nothing at all is even changed for the fermionic fields (\ref{electron}) and (\ref{neutrino}) as well. So all characteristic equations are of the simplest $n^{2}\equiv0$ form and therefore all solutions are of the light-like type alone, so that we can conclude that all fields of the standard model have causal propagation and what is more important is that this solution is ensured by a characteristic equation that is unchanged with respect to the one we had before the breakdown of the symmetry generating mass.

So far we have proved the causal propagation of all fields of the standard model as it has also been done in \cite{f-p}: however in this paper Farkas and Pocsik get for the Proca fields a characteristic equation given by $n^{2}\left(v+H\right)=0$ which is not correct; furthermore the authors disregard all other matter fields with no justification. In this paper we have obtained $n^{2}=0$ as the characteristic equation for all fields, which does not only tell us that causality is preserved but also that it is preserved because the propagation is not affected by the symmetry breaking for the mass generation.
\subsection{The Kibble-Sciama coupling}
Finally, we remind to the reader the Kibble-Sciama coupling prescription used to include the spin-torsion interaction \cite{k} and \cite{s}.

In the scheme of Einstein gravity as a spacetime metric the energy-curvature interaction is established through field equations coming from conservation laws in the form of the Jacobi-Bianchi identities; Kibble-Sciama adopted the same spirit in the most general case where spacetime torsion is allowed obtaining that the spin-torsion interaction is established through field equations coming from conservation laws in the form of the Jacobi-Bianchi identities written in the most general instance where torsion is considered: the resulting theory is the Einstein-Sciama-Kibble theory in which there is the energy-curvature as well as the spin-torsion coupling as basis for the field equations.

Now although all fields have energy and so all of them have an energy-curvature coupling, the scalar field has no spin and the gauge fields have spin but their spin tensor is force to vanish because of gauge invariance leaving only the gravitational and fermionic fields with a spin-torsion coupling; after torsion is separated and written in terms of the fermion fields themselves, we have that in the gravitational field equations additional potentials appear but it is in the fermionic field equations that these additional potentials have the features of autointeractions of the fermions with themselves and with one another: however as a closer inspection of field equations (\ref{electron}) and (\ref{neutrino}) may show we have that all torsional-spinorial contributions can actually be rearranged in order to be written either in the form of vector and scalar fields which are absorbed into the Proca and Higgs fields or in terms of pseudo-scalar fields which does not contain physical degrees of freedom, and therefore none of them is likely to influence any of the field equations in a phenomenologically detectable way.

Actually the idea of a strict connection relating gravitational extensions to the electroweak interactions has been suggested for instance in \cite{h}, \cite{n-l} and \cite{ds-s}: in these papers those authors use torsion or other tensors or connections to attempt a description that accounts for the electroweak forces or at least for their effects. Here we have proved that is possible to use generalized connections to mimic the electroweak force or at least its effects, but the problem is precisely that so long as the fundamental electroweak force is present, the two types of  derived or original electroweak forces do have the same structure, and their discrimination may be impossible; moreover when directly compared, it is likely that the derived one is negligible with respect to the original electroweak force, with the consequence that the former may always be swamped by the latter away. Then, although the idea of using the geometrical framework of general relativity to produce the electroweak-like dynamics is appealing and reasonable and may even be concrete, nevertheless there is the chance that we will never be able to observe such dynamics.
\section*{Conclusion}
In this paper we have proved the causal propagation of all fields of the standard model; moreover we have proved that causality is maintained because the entire propagation is unchanged during the generation of the masses of all massive particles in the standard model itself. Eventually we have discussed the spin-torsion coupling showing that all contributions may even influence the system of field equations but the possibility to actually observe such influence is unlikely.

\

\noindent \textbf{Acknowledgments.} I am grateful to Professor Giorgio Velo for valuable help.

\


\end{document}